\documentclass[runningheads]{llncs}
\usepackage{placeins}
\usepackage{graphicx}
\usepackage{amsmath}
\usepackage{amssymb}
\usepackage{amsfonts}
\usepackage{subcaption}
\begin{document}
\title{Introduction to StarNEig --- A Task-based Library for Solving Nonsymmetric Eigenvalue Problems}
\titlerunning{A Task-based Library for Solving Nonsymmetric Eigenvalue Problems}
%
\author{Mirko Myllykoski\orcidID{0000-0002-3689-0899} \and Carl~Christian~Kjelgaard~Mikkelsen\orcidID{0000-0002-9158-1941}}
\authorrunning{M. Myllykoski \and C.C Kjelgaard Mikkelsen}
%
\institute{Department of Computing Science, Ume{\aa} University, SE-901 87 Ume{\aa}, Sweden
\email{\{mirkom,spock\}@cs.umu.se}}
\maketitle              
\begin{abstract}
In this paper, we present the StarNEig library for solving dense non-symmetric (generalized) eigenvalue problems. 
The library is built on top of the StarPU runtime system and targets both shared and distributed memory machines. 
Some components of the library support GPUs.
The library is currently in an early beta state and only real arithmetic is supported.
Support for complex data types is planned for a future release.
This paper is aimed for potential users of the library.
We describe the design choices and capabilities of the library, and contrast them to existing software such as ScaLAPACK.
StarNEig implements a ScaLAPACK compatibility layer that should make it easy for a new user to transition to StarNEig.
We demonstrate the performance of the library with a small set of computational experiments.
\end{abstract}

\section{Introduction}

In this paper, we present the StarNEig library \cite{starneig} for solving dense non-symmetric (generalized) eigenvalue problems. 
StarNEig differs from the existing libraries such as LAPACK and ScaLAPACK in that it is relies on a modern task-based approach.
More specifically, StarNEig is build on top of the StarPU runtime system \cite{starpu}.
This allows StarNEig to target both shared memory and distributed memory machines. 
Furthermore, some components of the StarNEig library support GPUs.
The library is currently in an early beta state and under under continuous development.

This paper targets potential users of the library.
We hope that readers, who are already familiar with ScaLAPACK, will be able to decide if StarNEig is suitable for them.
In particular, we want to communicate what type of changes are necessary to make their software work with StarNEig.
We will explain, through an example, why the task-based approach can potentially lead to superior performance when compared to older, but well-established, approaches.
We also present a small sample of computational results which demonstrate the expected performance of the library.
We refer the reader to \cite{D27} for more comprehensive performance and accuracy evaluations.

The rest of this paper is organized as follows: 
Section \ref{sec:eigenproblem} provides a brief introduction to the solution of dense eigenvalue problems.
Section \ref{sec:task-based} explains the task-based approach and Section \ref{sec:starneig} introduces the reader to some of the inner workings of StarNEig.
Section \ref{sec:performance} presents a small set of computational results and, finally, Section \ref{sec:summary} concludes the paper.

\section{Solution of Dense Nonsymmetric Eigenvalue Problems} \label{sec:eigenproblem}

Given a matrix $A \in \mathbb{R}^{n \times n}$, the standard eigenvalue problem consists of computing eigenvalues $\lambda_i \in \mathbb{C}$ and matching eigenvectors $x_i \in \mathbb{C}^n$ such that
\begin{align}
 A x_i = \lambda_i x_i, \quad x_i \not = 0.
\end{align}
Similarly, given matrices $A \in \mathbb{R}^{n \times n}$ and $B \in \mathbb{R}^{n \times n}$ the generalized eigenvalue problem for the matrix pair $(A,B)$ consists of computing generalized eigenvalues $\lambda_i \in \mathbb{C}$ and matching generalized eigenvectors $x_i \in \mathbb{C}^n $ such that
\begin{align}
 A x_i = \lambda_i B x_i, \quad x_i \not = 0.
\end{align}

If the matrix $A$ or the matrices $A$ and $B$ are dense and nonsymmetric, then route of acquiring the (generalized) eigenvalues and the (generalized) eigenvectors usually includes the following three steps:
\begin{description}

 \item[Hessenberg(-triangular) reduction:] The matrix $A$ or the matrix pair $(A,B)$ is reduced to upper Hessenberg or Hessenberg-triangular form by an orthogonal similarity transformation
 \begin{align}
  A = Q_1 H Q_1^T \; \text{ or } \; (A,B) = Q_1 (H,R) Z_1^T,
 \end{align}
 where $H$ is upper Hessenberg, $R$ is a upper triangular, and $Q_1$ and $Z_1$ are orthogonal.
 
 \item[Schur reduction:] The Hessenberg matrix $H$ or the Hessenberg-triangular matrix pair $(H,R)$ is reduced to Schur or generalized Schur form  by an orthogonal similarity transformation
 \begin{align}
  H = Q_2 S Q_2^T \; \text{ or } \; (H,R) = Q_2 (S,T) Z_2^T,
 \end{align}
 where $S$ is upper quasi-triangular with $1 \times 1$ and $2 \times 2$ blocks on the diagonal, $T$ is a upper triangular, and $Q_2$ and $Z_2$ are orthogonal.
 The eigenvalues or generalized eigenvalues can be determined from the diagonal blocks of $S$ or $(S,T)$.
 
 \item[Eigenvectors:] Finally, we compute vectors $y_i \in \mathbb{C}^n$ from
 \begin{align}
  (S - \lambda_i I) y_i = 0 \; \text{ or } \; (S - \lambda_i T) y_i = 0
 \end{align}
 and backtransform to the original basis by
 \begin{align}
  x_i = Q_1 Q_2 y_i \; \text{ or } \; x_i = Z_1 Z_2 y_i.
 \end{align}
 
\end{description}
Additionally, a fourth step can be performed to acquire a desired invariant subspace of $A$ or $(A,B)$:
\begin{description}
 \item[Eigenvalue reordering:] The Schur form $S$ or the generalized Schur form $(S,T)$ is reordered, such that a selected set of eigenvalues or generalized eigenvalues appears in the leading diagonal blocks of an updated Schur form $\hat S$ or an updated generalized Schur form $(\hat S, \hat T)$, by an orthogonal similarity transformation
 \begin{align}
  S = Q_3 \hat S Q_3^T \; \text{ or } \; (S,T) = Q_3 (\hat S, \hat T) Z_3^T,
 \end{align}
 where $Q_3$ and $Z_3$ are orthogonal.
\end{description}
See \cite{Golub1996} for a detailed explanation of the underlying mathematical theory.

\section{A Case for the Task-Based Approach} \label{sec:task-based}

A task-based algorithm functions by cutting the computational work into self-contained tasks that all have a well defined set of inputs and outputs.
The algorithm inserts the tasks into a runtime system that derives the task dependences and schedules the tasks to computational resources in a sequentially consistent order.
As long as the cutting is carefully done, the underlying parallelism is exposed automatically as the runtime system unravels the resulting task graph.

\subsection{Novelty in StarNEig}

The first main source of novelty in StarNEig comes from the way in which the computational work is cut into tasks.
As with many other task-based matrix algorithms, the matrices are divided into (square) tiles and each task takes a set of tiles as its input and produces/modifies a set of tiles as its output.
The Hessenberg reduction, Schur reduction and eigenvalue reordering steps are based on two-sided transformation algorithms.
These algorithms lead to data dependences that are much more complicated than the dependences arising from one-sided transformation algorithm such as the LU factorization.
The second main source of novelty in StarNEig is related to the eigenvector computation step. 
Here, the data dependences are comparatively simple but the computations must be protected against floating-point overflow.
This is a nontrivial issue to address in a parallel setting; see \cite{ppam2017cckm,ppam2019cckm,ccpe2018cckm}.

Furthermore, the Schur reduction and eigenvalue reordering steps apply a series of overlapping local transformations to the matrices.
Due to this overlap, the two computational steps cannot have have a clear one-to-one mapping between the tasks and the (output) tiles since the local transformations must at some point cross between two or more tiles.
Instead, most task ends up modifying several tiles and this can introduce spurious data dependences\footnote{A spurious data dependency is created when two (or more) tasks modify non-overlapping parts of the same tile but the runtime system interprets this as a true data dependency.} that limit the concurrency. 

\subsection{Bulge Chasing and Eigenvalue Reordering}

We will now use the Schur reduction and eigenvalue reordering steps to illustrate some benefits of the task-based approach.
The modern approach for obtaining a Schur form $S$ of $A$ is to apply the multishift QR algorithm with Aggressive Early Deflation (AED) to the upper Hessenberg form $H$ (see \cite{GraKagKreShao2015b} and references therein).
The algorithm is a sequence of steps of two types: AED and bulge chasing. 
The bulge chasing step creates a set of bulges which are chased down the diagonal to complete one pipelined QR iteration.
This is accomplished by applying sequences of overlapping $3 \times 3$ Householder reflectors to $H$.
Similarly, the eigenvalue reordering step is based on applying sequences of overlapping Givens rotations and $3 \times 3$ Householder reflectors to $S$.

\begin{figure}[t]
\centering
\begin{subfigure}[t]{0.3\textwidth}
 \includegraphics[width=\textwidth]{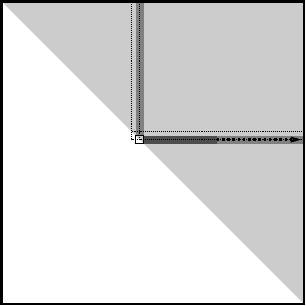}
 \caption{Scalar updates.}
 \label{fig:differences_a}
\end{subfigure}
\;
\begin{subfigure}[t]{0.3\textwidth}
 \includegraphics[width=\textwidth]{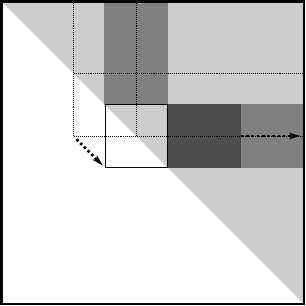}
 \caption{Accumulated updates.}
 \label{fig:differences_b}
\end{subfigure}

\begin{subfigure}[t]{0.3\textwidth}
 \includegraphics[width=\textwidth]{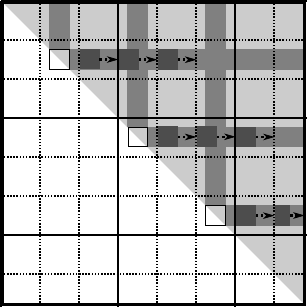}
 \caption{ScaLAPACK.}
 \label{fig:differences_c}
\end{subfigure}
\;
\begin{subfigure}[t]{0.3\textwidth}
 \includegraphics[width=\textwidth]{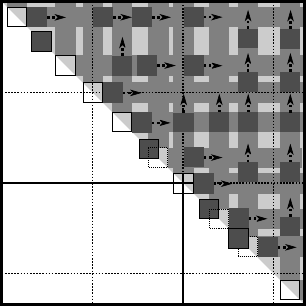}
 \caption{StarNEig.}
 \label{fig:differences_d}
\end{subfigure}
\caption{ 
 Hypothetical snapshots taken during the computations.
 Active regions are highlighted with darker shade and the propagation directions of the transformations are marked with arrows.
 In (a) and (b), the overlap between two (accumulated) transformations is highlighted with dashed lines. 
 In (c) and (d), the dashed lines illustrate how the matrix is divided into distributed blocks and the solid lines illustrate the MPI process mesh.
}\label{fig:differences}
\end{figure}

If the local transformations are applied one by one, then memory is accessed as shown in Fig. \ref{fig:differences_a}.
This is grossly inefficient for two reasons: i) the transformation is so localized that parallelizing it would not produce any speedup and ii) the matrix elements are touched only once thus leading to very low arithmetic intensity.
The modern approach groups a set of local transformation together and initially applies them to a relatively small diagonal window as shown in Fig. \ref{fig:differences_b}.
The local transformations are accumulated into an accumulator matrix and later applied as level-3 BLAS operations acting on the relevant sections of the matrix.
This leads to much higher arithmetic intensity and enables proper parallel implementations as \textit{multiple} diagonal windows can be processed simultanously.

In particular, the Schur reduction and eigenvalue reordering steps are implemented in ScaLAPACK as PDHSEQR \cite{GraKagKreShao2015b} and PDTRSEN \cite{Granat_2009} subroutines, receptively.
Following a typical ScaLAPACK formula, the matrices are distributed in two-dimensional block cyclic fashion.
The resulting memory access pattern is illustrated in Fig. \ref{fig:differences_c} for a $3 \times 3$ MPI process mesh.
In this particular example, three diagonal windows can be processed simultaneously. 
The related level-3 BLAS updates require careful coordination since the left and right hand side updates must be performed in a sequentially consistent order.
In practice, this means (global or broadcast) synchronization after each set of (left or right hand side) updates have been applied.

\begin{figure}[t]
 \centering
 \includegraphics[scale=0.4]{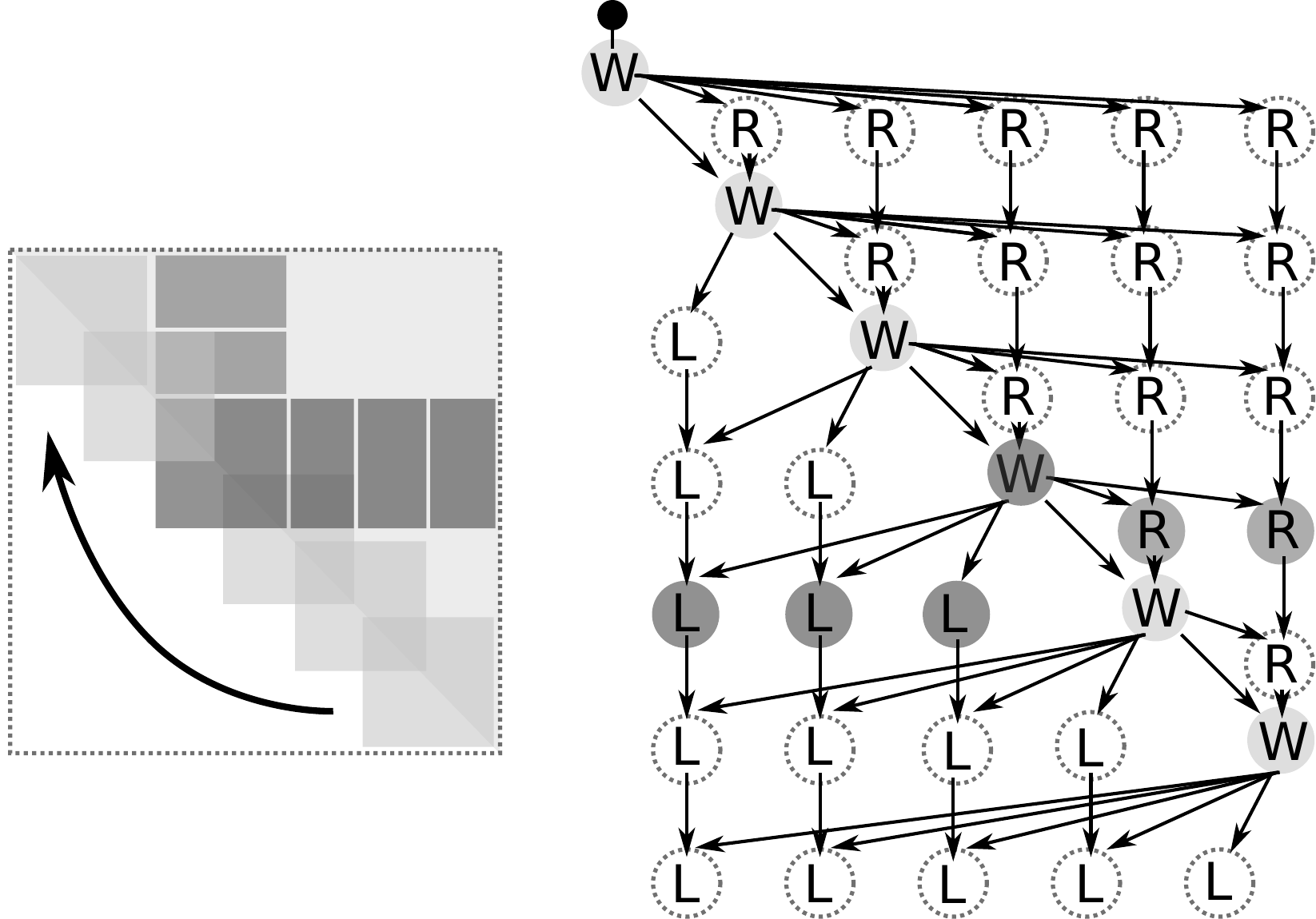}
 \caption{
 A hypothetical task graph arising from a situation where a Schur form is reordered with a single chain of diagonal windows.
 We have simplified the graph by leaving out the dependences between the right (R) and left (L) update tasks as these dependences are already enforced through the window tasks (W).
 }
 \label{fig:chain_flow}
\end{figure}

In a task-based approach, this can be done using the following task types:
\begin{description}
 \item[Window task] applies a set of local transformations inside the diagonal window. Takes the intersecting tiles as input, and produces updated tiles and an accumulator matrix as output. 
 \item[Right update task] applies accumulated right-hand side updates using level-3 BLAS operations. Takes the intersecting tiles and an accumulator matrix as input, and produces updated tiles as output.
 \item[Left update task] applies accumulated left-hand side updates using level-3 BLAS operations. Takes the intersecting tiles and an accumulator matrix as input, and produces updated tiles as output.
\end{description}

The tasks are inserted in a sequentially consistent order and each window chain leads to a task graph like the one shown in Fig. \ref{fig:chain_flow}.
It is critical to realize that the runtime system \textit{guarantees} that the tasks are executed in a sequentially consistent order.
In particular, there is no need for synchronization and different stages are allowed to overlap as illustrated in Fig. \ref{fig:differences_d}.
This leads to much higher concurrency.
Under suitable conditions, the AED step can also be partially overlapped with the bulge chasing step.
Other benefits of the task-based approach include, for example, better load balancing, task priorities, accelerators support and implicit MPI communication.
See \cite{Myllykoski2018,NLAFET-WN11,D26,D27} for implementation details and further information.

\section{StarNEig Library} \label{sec:starneig}

StarNEig is a C-library that runs on top of the StarPU task-based runtime system.
StarPU handles low-level operations such as heterogeneous scheduling, data transfers and replication between various memory spaces, and MPI communication between compute nodes.
In particular, StarPU is responsible for managing the various computational resources such as CPU cores and GPUs.
In order to accomplish this, StarPU creates a set of worker threads; usually one thread per computational resource.
In addition, one thread is responsible for inserting the tasks and tracking the state of the machine. 
If necessary, one additional thread is allocated for MPI communication.
Thus, StarNEig should be used in a \textit{one process per node} (1ppn) configuration, i.e., several CPU cores should be allocated for each process (a node can be a full compute node, a NUMA island or some other collection of CPU cores).

\begin{table}[t]
\caption{
 Current status of the StarNEig library.
 }
\label{tab:starneig_status}
\centering
\begin{tabular}{l|ccc}
\textbf{Step}         & \textbf{Shared memory} & \textbf{Distr. memory} & \textbf{GPUs} \\ \hline
\textbf{Hessenberg}   & Complete               & ScaLAPACK                   & Single GPU    \\
\textbf{Schur}        & Complete               & Complete                    & Experimental  \\
\textbf{Reordering }  & Complete               & Complete                    & Experimental  \\
\textbf{Eigenvectors} & Complete               & In progress                 & Not planned   \\ \hline
\textbf{Hessenberg-triangular}    & Planned (LAPACK)       & ScaLAPACK                   & Not planned   \\
\textbf{Generalized Schur}        & Complete               & Complete                    & Experimental  \\
\textbf{Generalized reordering}   & Complete               & Complete                    & Experimental  \\
\textbf{Generalized eigenvectors} & Complete               & In progress                 & Not planned
\end{tabular}
\end{table}

The current status of the StarNEig library is summarized in Table \ref{tab:starneig_status}.
The library is currently in an early beta state.
At the time of writing this paper, only real arithmetic is supported and certain interface functions are implemented as LAPACK and ScaLAPACK wrapper functions.
The library is under continuous development.
In particular, additional distributed memory functionality and support for complex data types are planned for a future release.

\subsection{Distributed Memory}

\begin{figure}[t]
\centering
\begin{subfigure}[t]{0.3\textwidth}
 \includegraphics[width=\textwidth]{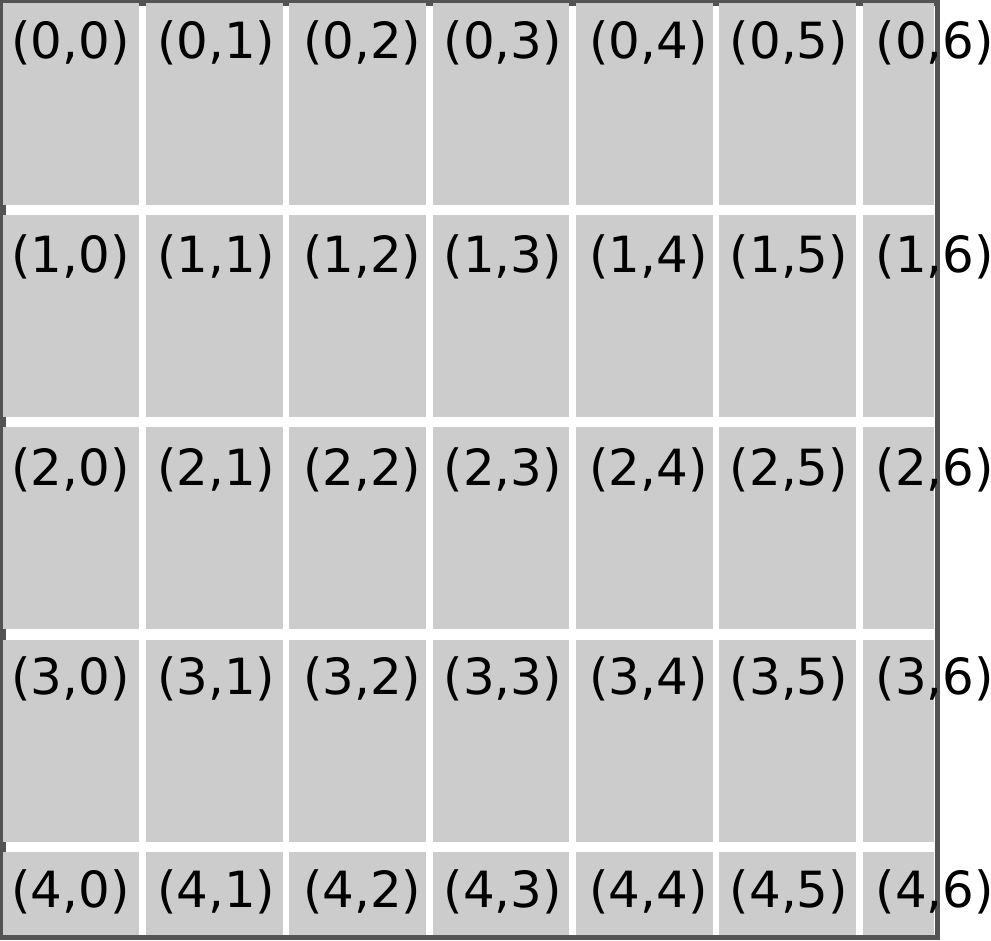}
 \caption{Distributed blocks.}
 \label{fig:distr_matrix1}
\end{subfigure}
\begin{subfigure}[t]{0.3\textwidth}
 \includegraphics[width=\textwidth]{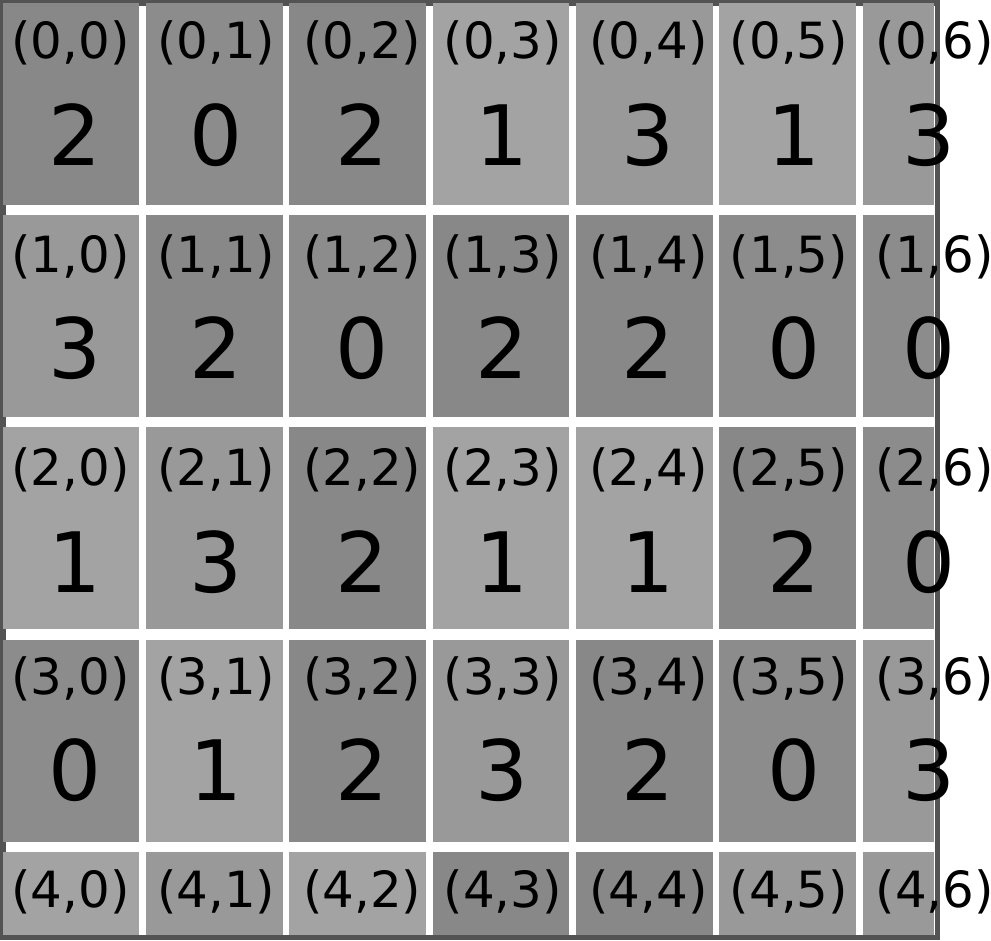}
 \caption{Arbitrary mapping.}
 \label{fig:distr_matrix2}
\end{subfigure}
\begin{subfigure}[t]{0.3\textwidth}
 \includegraphics[width=\textwidth]{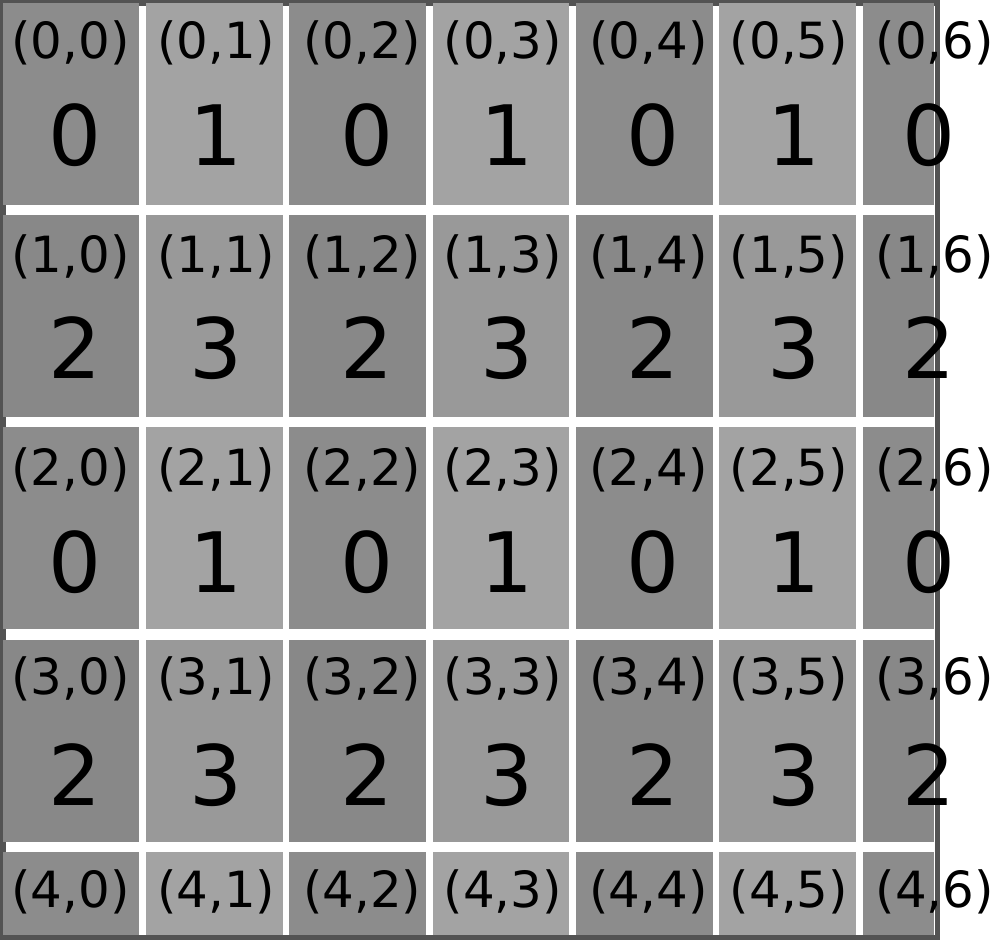}
 \caption{2D-BCD mapping.}
 \label{fig:distr_matrix3}
\end{subfigure}
\caption{Examples of various data distributions supported by StarNEig, including two-dimensional block cyclic distribution (2D-BCD).}\label{fig:distr_matrix}
\end{figure}

StarNEig distributes the matrices in rectangular blocks of a uniform size (excluding the last block row and column) as illustrated in Fig. \ref{fig:distr_matrix1}.
The data distribution, i.e., the mapping from the distributed blocks to the MPI process rank space, can be arbitrary as illustrated in Fig. \ref{fig:distr_matrix2}.
A user has three options:
\begin{enumerate}
 \item Use the default data distribution. 
 This is recommended for most users and leads to reasonable performance in many cases.
 \item Use a two-dimensional block cyclic distribution (see Fig. \ref{fig:distr_matrix3}). In this case, the user may select the MPI process mesh dimensions and the rank ordering. 
 \item Define a data distribution function $d : \mathbb{Z^+} \times \mathbb{Z^+} \to \mathbb{Z^+}$ that maps the block row and column indices to the MPI rank space.
 For example, in Fig. \ref{fig:distr_matrix2}, the rank 0 owns the blocks (0,1), (1,2), (1,5), (1,6), (2,6), (3,0) and (3,5).
\end{enumerate}
The library implements distribution agnostic copy, scatter and gather operations.

Users who are familiar with ScaLAPACK are likely accustomed to using relatively small distributed block sizes (between 64--256).
In contrast, StarNEig functions optimally only if the distributed blocks are relatively large (at least 1000).
This is due to the fact that StarNEig further divides the distributed blocks into tiles and a tiny tile size leads to excessive task scheduling overhead because the tile size is closely connected to the task granularity.
Furthermore, as mentioned in the preceding section, StarNEig should be used in 1ppn configuration as opposed to a \textit{one process per core} (1ppc) configuration which is more common with ScaLAPACK.

\subsection{ScaLAPACK Compatibility}

StarNEig is fully compatible with ScaLAPACK and provides a ScaLAPACK compatibility layer that encapsulates BLACS contexts and descriptors \cite{blacs} inside transparent objects, and implements a set of bidirectional conversion functions.
The conversions are performed in-place and do not modify any of the underlying data structures.
Thus, users can mix StarNEig interface functions with ScaLAPACK subroutines without intermediate conversions.

\section{Performance Evaluation} \label{sec:performance}

\begin{table}[p]
\centering
\caption{
 A run time comparison between ScaLAPACK and StarNEig.
 }
\label{tab:sep_schur_compare}
\begin{tabular}{r | c c | c c | c c}
                         & \multicolumn{2}{c|}{CPU cores} & \multicolumn{2}{c|}{Schur reduction \textit{(secs)}} & \multicolumn{2}{c}{Eigenvalue reordering \textit{(secs)}} \\
\multicolumn{1}{c|}{$n$} & \;ScaLAPACK\; & \;StarNEig\;     & \;PDHSEQR\; & \;StarNEig\; & \;PDTRSEN\; & \;StarNEig\; \\ \hline
10\,000 & 36 & 28 & 38 & 18 & 12 & 3 \\
20\,000 & 36 & 28 & 158 & 85 & 72 & 25 \\
40\,000 & 36 & 28 & 708 & 431 & 512 & 180 \\
60\,000 & 121 & 112 & 992 & 563 & 669 & 168 \\
80\,000 & 121 & 112 & 1667 & 904 & 1709 & 391 \\
100\,000 & 121 & 112 & 3319 & 1168 & 3285 & 737 \\
120\,000 & 256 & 252 & 3268 & 1111 & 2902 & 581 \\

\end{tabular}
\end{table}

\begin{figure}[p]
 \centering
 \includegraphics[scale=0.7]{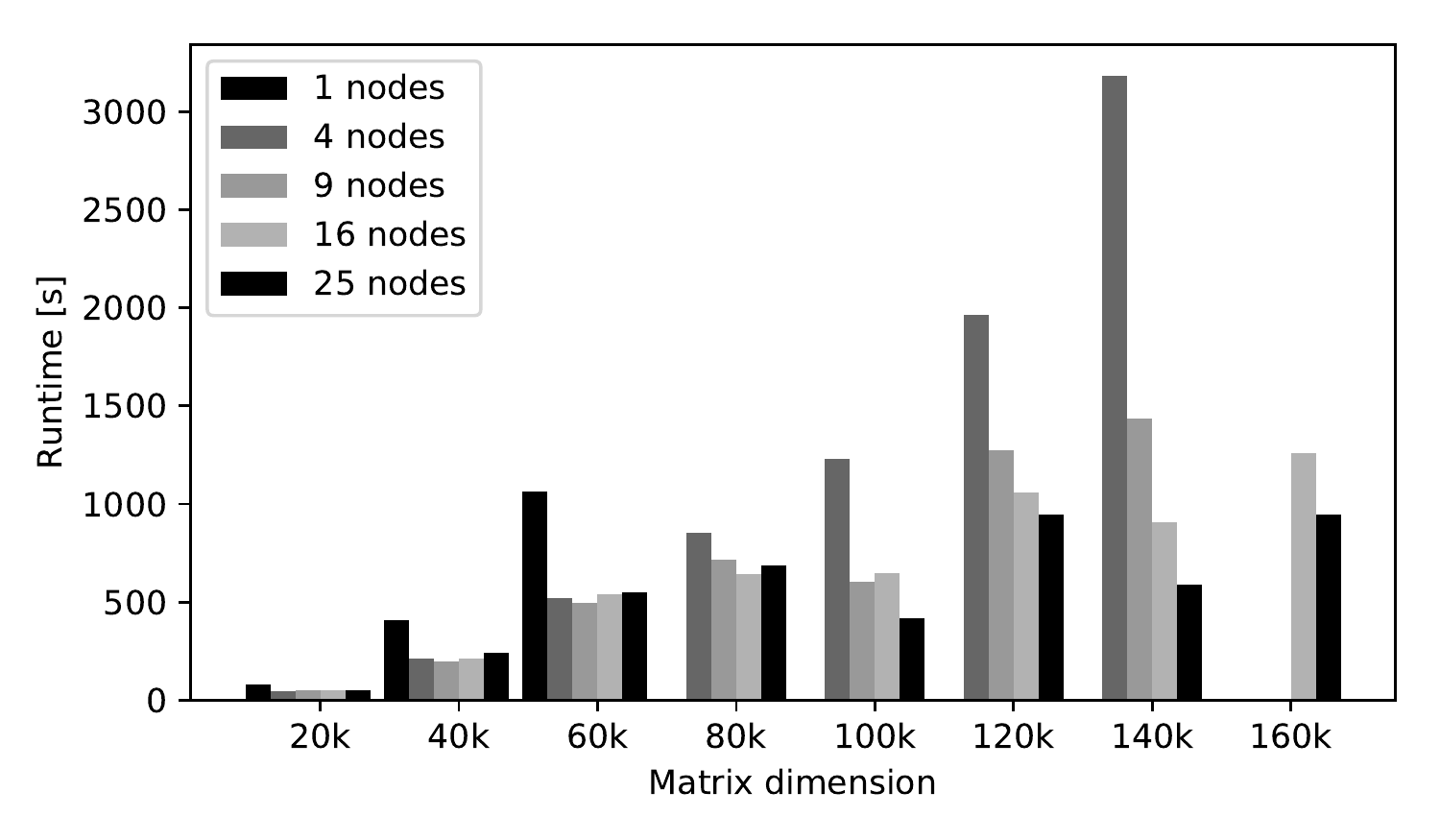}
 \includegraphics[scale=0.7]{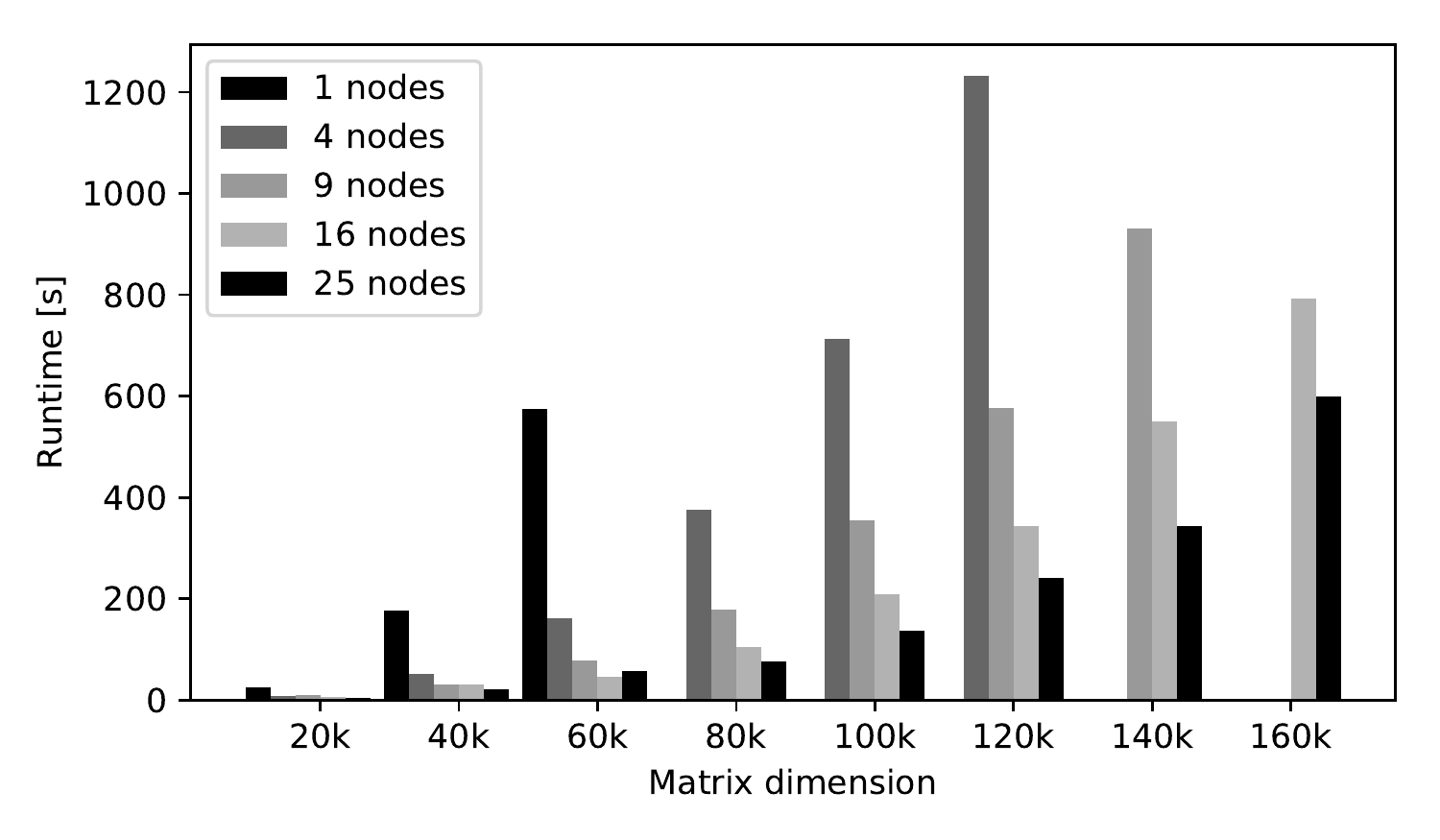}
 \caption{
 Distributed memory scalability when computing a Schur form (upper figure) and reordering a Schur form (bottom figure).
 Each node contains 28 CPU cores.
 }
 \label{fig:sep_compare}
\end{figure}

All computational experiments were performed on the Kebnekaise system, located at the High Performance Computing Center North (HPC2N), Ume{\aa} University.
Each compute node contains 28 Intel Xeon E5-2690v4 cores (2 NUMA islands with 14 cores each) and 128 GB memory. 
The nodes are connected with FDR Infiniband.
The software was compiled with GCC 7.3.0 and linked to OpenMPI 3.1.3, OpenBLAS 0.3.2, ScaLAPACK 2.0.2, and StarPU 1.2.8. 
All experiments were performed using a square MPI process grid.
We always map each StarNEig process to a full node (28 cores) and each ScaLAPACK process to a single CPU core.
The number of CPU cores in each ScaLAPACK experiment is always equal or larger than the number of CPU cores in the corresponding StarNEig experiment.
The upper Hessenberg matrices for the Schur reduction experiments were computed from random matrices (entries uniformly distributed over the interval $[-1,1]$).

Table \ref{tab:sep_schur_compare} shows a comparison between ScaLAPACK and StarNEig.
We note that StarNEig is between 1.6 and 2.9 times faster than PDHSEQR and between 2.8 and 5.0 times faster than PDTRSEN.
Figure \ref{fig:sep_compare} gives some idea of how well the library is expected to scale in distributed memory.
We note that StarNEig scales reasonably when computing the Schur form and almost linearly when reordering the Schur form.
The iterative nature of the QR algorithm makes the Schur reduction results less predictable.
See \cite{NLAFET-WN11,D27} for more comprehensive comparisons.

\section{Summary} \label{sec:summary}

This paper presented a new library called StarNEig.
The paper is aimed for potential users of the library.
Various design choices were explained and contrasted to existing software.
In particular, users who are already familiar with ScaLAPACK should know following:
\begin{itemize}
 \item StarNEig expect that the matrices are distributed in relatively large blocks compared to ScaLAPACK.
 \item StarNEig should be used in a \textit{one process per node} (1ppn) configuration as opposed to a \textit{one process per core} (1ppc) configuration which is very common with ScaLAPACK.
 \item StarNEig implements a ScaLAPACK compatibility layer.
\end{itemize}
The presented distributed memory results indicate that the library is highly competitive with ScaLAPACK.
The authors hope to start a discussion which would help guide and prioritize the future development of the library.

\section*{Acknowledgements} 
\label{sec:acknowledgements}

StarNEig has been developed by the authors, Angelika Schwarz (who has written the standard eigenvector solver), Lars Karlsson, and Bo K{\aa}gstr{\"o}m.
This work is part of a project (NLAFET) that has received funding from the European Union's Horizon 2020 research and innovation programme under grant agreement No 671633. This work was supported by the Swedish strategic research programme eSSENCE. We thank the High Performance Computing Center North (HPC2N) at Ume{\aa} University for providing computational resources and valuable support during test and performance runs.

%
%
%
%
\bibliographystyle{splncs04}
\bibliography{myllykoski}
\end{document}